\documentclass[fleqn,12pt,twoside]{article}
\usepackage{espcrc1}
\usepackage{epsfig}

\usepackage{graphicx}
\usepackage[figuresright]{rotating}

\newcommand{\AmS}{{\protect\the\textfont2
  A\kern-.1667em\lower.5ex\hbox{M}\kern-.125emS}}

\title{Flow and Bose-Einstein Correlations in Au-Au Collisions at RHIC}

\author {Steven Manly for the PHOBOS Collaboration \\ \mbox{  } \\
B.B.Back$^1$, M.D.Baker$^2$, D.S.Barton$^2$, R.R.Betts$^6$,
R.Bindel$^7$, A.Budzanowski$^3$, W.Busza$^4$, A.Carroll$^2$,
M.P.Decowski$^4$, E.Garcia$^6$, N.George$^1$, K.Gulbrandsen$^4$,
S.Gushue$^2$, C.Halliwell$^6$, J.Hamblen$^8$, C.Henderson$^4$,
D.Hofman$^6$, R.S.Hollis$^6$, R.Ho\l y\'{n}ski$^3$, B.Holzman$^2$,
A.Iordanova$^6$, E.Johnson$^8$, J.Kane$^4$, J.Katzy$^{4,6}$,
N.Khan$^8$, W.Kucewicz$^6$, P.Kulinich$^4$, C.M.Kuo$^5$,
W.T.Lin$^5$, S.Manly$^{8}$,  D.McLeod$^6$, J.Micha\l owski$^3$,
A.Mignerey$^7$, R.Nouicer$^6$, A.Olszewski$^{3}$, R.Pak$^2$,
I.C.Park$^8$, H.Pernegger$^4$, C.Reed$^4$, L.P.Remsberg$^2$,
M.Reuter$^6$, C.Roland$^4$, G.Roland$^4$, L.Rosenberg$^4$, J.
Sagerer$^6$, P.Sarin$^4$, P.Sawicki$^3$, W.Skulski$^8$,
S.G.Steadman$^4$, P.Steinberg$^2$, G.S.F.Stephans$^4$,
M.Stodulski$^3$, A.Sukhanov$^2$, J.-L.Tang$^5$, R.Teng$^8$,
A.Trzupek$^3$, C.Vale$^4$, G.J.van Nieuwenhuizen$^4$,
R.Verdier$^4$, B.Wadsworth$^4$, F.L.H.Wolfs$^8$, B.Wosiek$^3$,
K.Wo\'{z}niak$^{3}$, A.H.Wuosmaa$^1$, B.Wys\l ouch$^4$\\ \mbox{
}\\ {\footnotesize $^1$ Argonne National Laboratory, $^2$
Brookhaven National Laboratory, $^3$ Institute of Nuclear Physics,
Krak\'{o}w, Poland, $^4$ Massachusetts Institute of Technology,
$^5$ National Central University, Chung-Li, Taiwan, $^6$
University of Illinois at Chicago, $^7$ University of Maryland,
$^8$ University of Rochester} }

\begin{document}

\maketitle

\begin{abstract}
Elliptic flow and Bose-Einstein correlations have been measured in
Au-Au collisions at $\sqrt{s_{_{NN}}} =$ 130 and 200 GeV using the
PHOBOS detector at RHIC. The systematic dependencies of the flow
signal on the transverse momentum, pseudorapidity, and centrality
of the collision, as well as the beam energy are shown.
In addition, results of a 3-dimensional analysis of two-pion
correlations in the 200 GeV data are presented.
\end{abstract}

\section{Introduction}

The evolution of the space-time structure of the particle emitting
source created in heavy ion collisions can be probed by measuring
the azimuthal anisotropy (e.g., elliptic flow) and two-particle
interferometry in the final state particle distributions. Elliptic
flow is thought to provide information on the early stages of the
collision, the nuclear equation of state and the degree of
equilibration attained during the evolution of the collision
\cite{r_r_review}, while two-particle interferometry provides
information on the temporal and spatial extent of the source
\cite{heinz}.

The results presented here are based on data taken during the
first two RHIC physics runs for Au-Au collisions at
$\sqrt{s_{_{NN}}} =$ 130 and 200 GeV. All 200 GeV results
presented here are preliminary. The PHOBOS detector employs
silicon pad detectors to perform tracking, vertex detection and
multiplicity measurements. Details of the setup and the layout of
the silicon sensors can be found elsewhere \cite{phobos2,phobos3}.
Event triggering and the determination of the centrality were
based on the information provided by two sets of scintillating
paddle counters \cite{phobosprl}. The raw data for these analyses
came in the form of energy depositions from the passage of charged
particles through individual detector pads, known as hits. The hit
definition and signal processing procedure used for the flow
analysis is described in reference \cite{flow130}.
The position of the primary collision vertex was determined on an
event-by-event basis by extrapolating tracks found in the
spectrometer arms and/or the vertex detector.
The event plane was determined by a standard subevent technique
\cite{pandv} using hits in symmetric and uniform regions of the
octagonal multiplicity detector. Charged particle tracks were
reconstructed in the spectrometer arms using techniques described
previously \cite{tracking1,tracking2}. Pions were identified using
the measured momentum and the specific ionization loss observed in
the spectrometer silicon detectors.

\section{Flow}

 Flow results from two
independent analyses are presented here.  In the first, known as
the hit-based analysis, the event plane was determined from hits
in the single layer Si multiplicity detectors and the second
Fourier coefficient of the hit azimuthal angle distribution (also
known as the elliptic flow), v$_{2}$, was evaluated by correlating
the event plane to hits in a different region of the multiplicity
detectors. In the second flow analysis, known as the track-based
analysis, v$_{2}$ was determined by correlating the event plane to
tracks found in the spectrometer arms.

\begin{figure}[htb]
\begin{minipage}[t]{80mm}
\centerline{ \epsfig{file=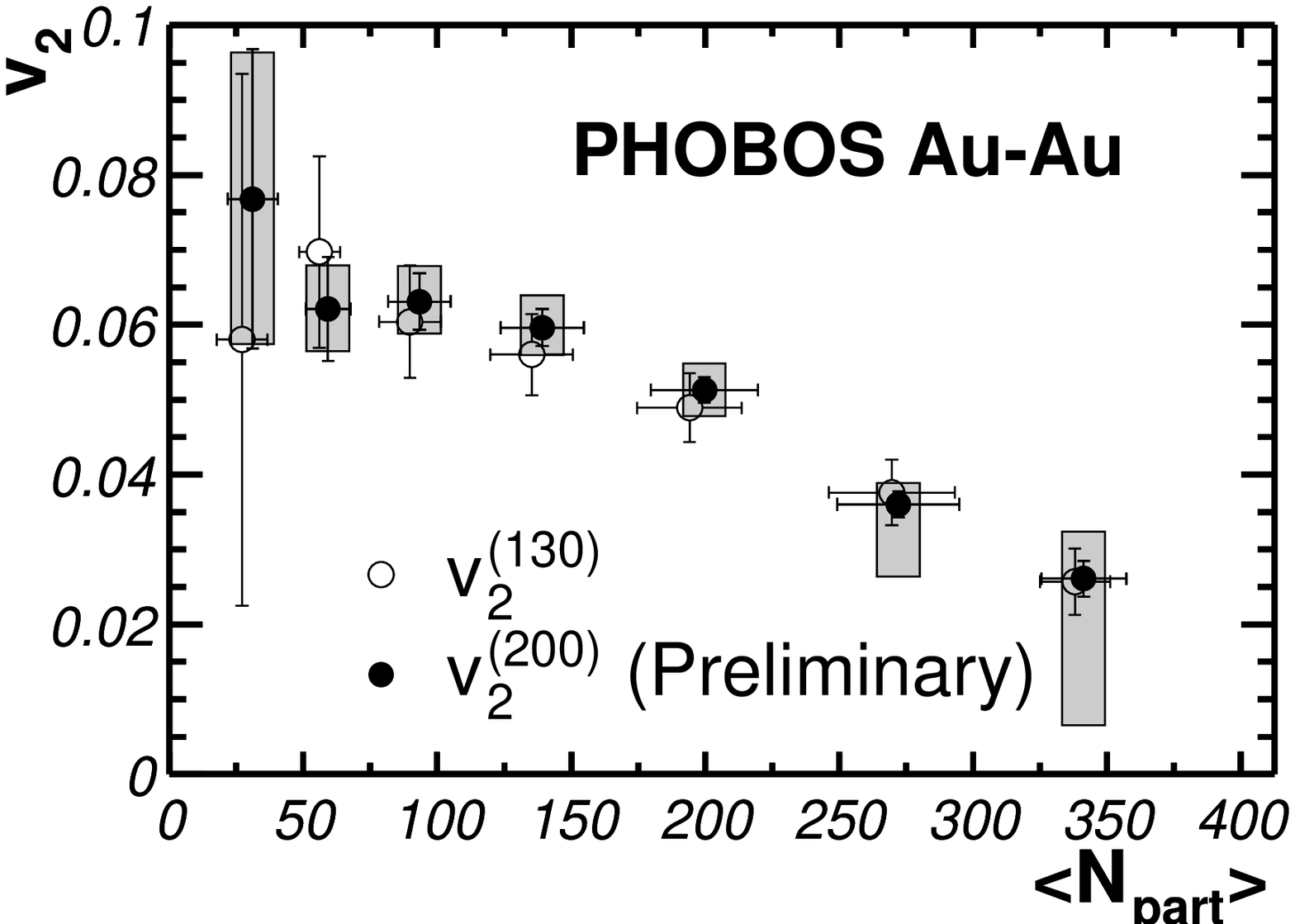,width=8cm} }
\caption{Elliptic flow as a function of the number of participants
for Au-Au collisions at $\sqrt{s_{_{NN}}}$ = 130 and 200 GeV.}
\label{v2npartener}
\end{minipage}
\hspace{\fill}
\begin{minipage}[t]{75mm}
\centerline{ \epsfig{file=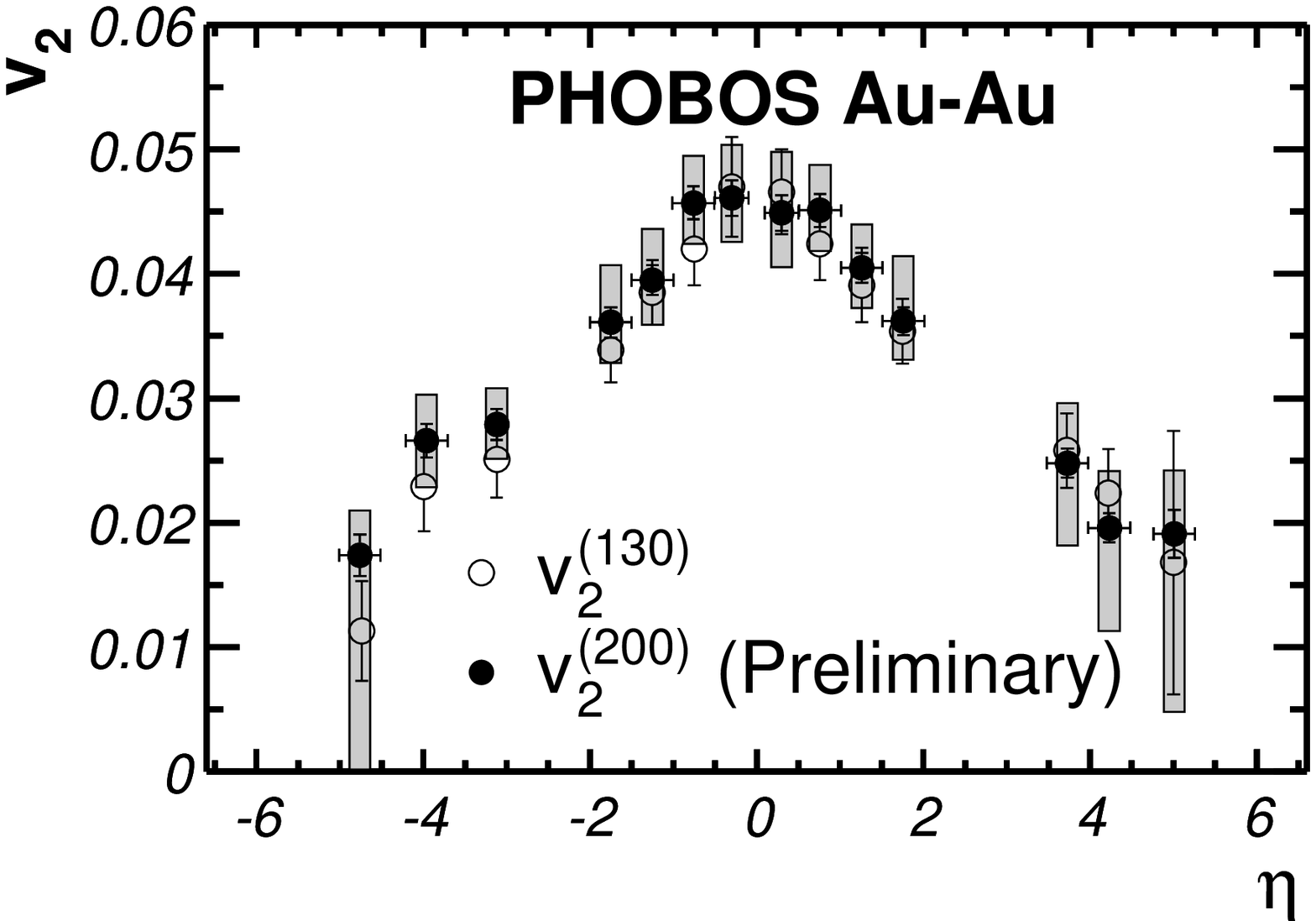,width=8cm} }
\caption{Elliptic flow as a function of pseudorapidity for Au-Au
collisions at $\sqrt{s_{_{NN}}}$ = 130 and 200 GeV.}
\label{v2etaener}
\end{minipage}
\end{figure}

For the hit-based analysis, events were chosen in a fiducial
region that maximized the $\eta$ coverage and event plane
sensitivity for the analysis. Equal multiplicity subevents were
defined in the regions $0.1<|\eta|<2$ for the event plane
determination and evaluation of the event plane resolution.
Details of this analysis are provided in reference \cite{flow130}.
Results from the hit-based flow analysis are shown in
Figures~\ref{v2npartener} and~\ref{v2etaener}.  The 1$\sigma$
statistical errors are shown for both analyses.  The 90\%
confidence level systematic errors are shown as boxes for the 200
GeV data points. As can be seen in these two figures, the flow
signal is little changed with the increase in the center-of-mass
energy of the collision from 130 to 200 GeV.  The unique, and very
nearly complete, $\eta$ coverage shown in Figure~\ref{v2etaener}
shows a substantial drop in v$_{2}$ as a function of $|\eta|$ that
is not yet understood \cite{flow130}.

In the track-based analysis, events were chosen in a fiducial
region that maximized the tracking efficiency and the reaction
plane sensitivity. Subevents were defined roughly in the regions
$2<|\eta|<3$ for the event plane determination and the evaluation
of the reaction plane resolution. Charged tracks reconstructed in
the spectrometer arms in the region $0<\eta<2.5$ were used to
determine the elliptic flow. In detail, the track-based flow
analysis is quite different from our previously released flow
analysis.  Differences include the use of a vertex dependent
reaction plane weighting and resolution correction, as well as a
larger separation of the subevents in $\eta$.
The flow signal is determined as the asymmetry in the track
azimuthal angle distribution measured relative to the event plane.
The track-based flow results are less dependent on Monte Carlo
corrections and less sensitive to background and non-flow
correlations as compared to the hit-based flow measurements.

\begin{figure}[htb]
\begin{minipage}[t]{80mm}
\centerline{ \epsfig{file=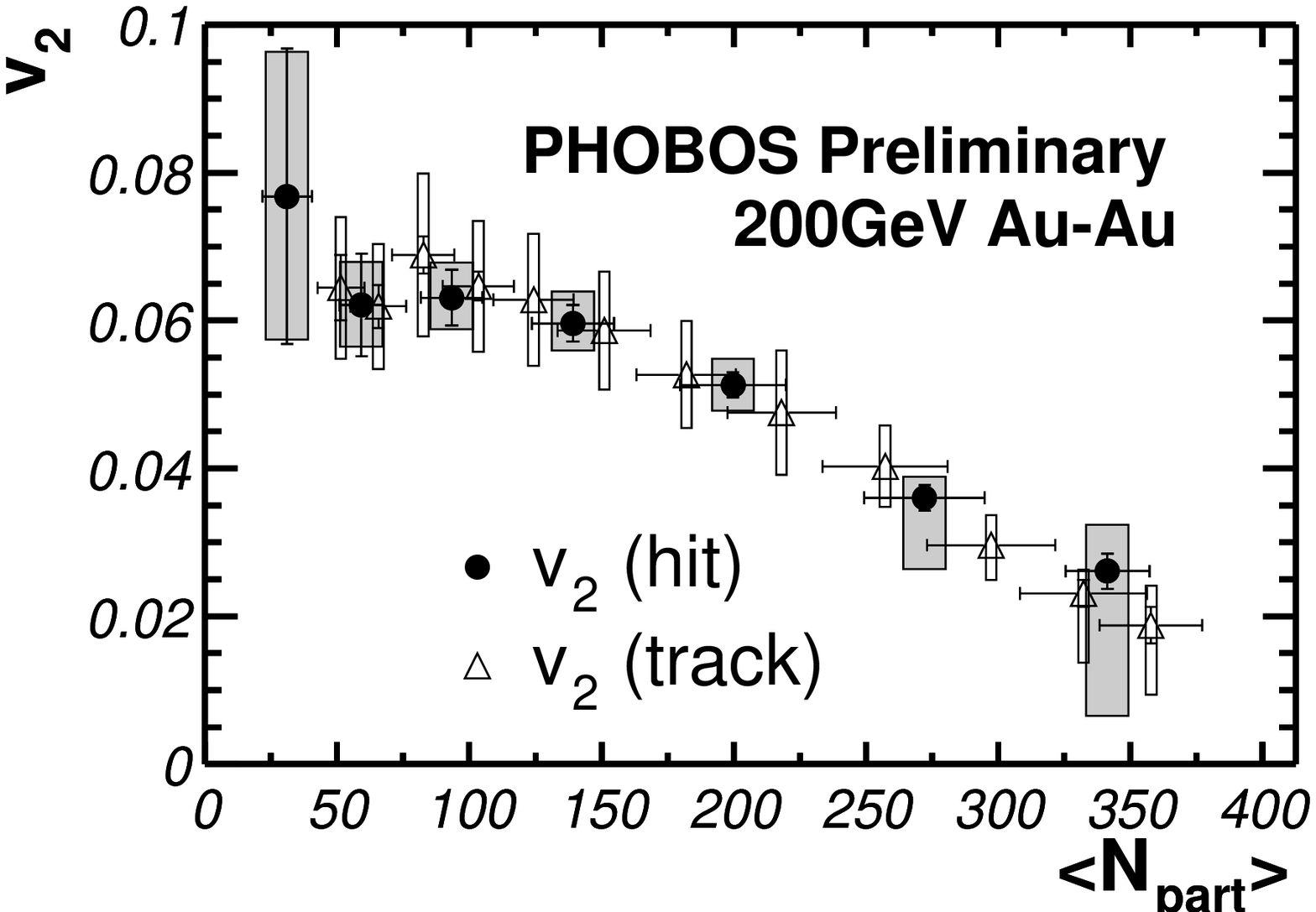,width=8cm} }
\caption{Elliptic flow as a function of the number of participants
for Au-Au collisions at $\sqrt{s_{_{NN}}}$ = 200 GeV for the
hit-based and track-based analyses.} \label{v2CenHitTrk}
\end{minipage}
\hspace{\fill}
\begin{minipage}[t]{75mm}
\centerline{ \epsfig{file=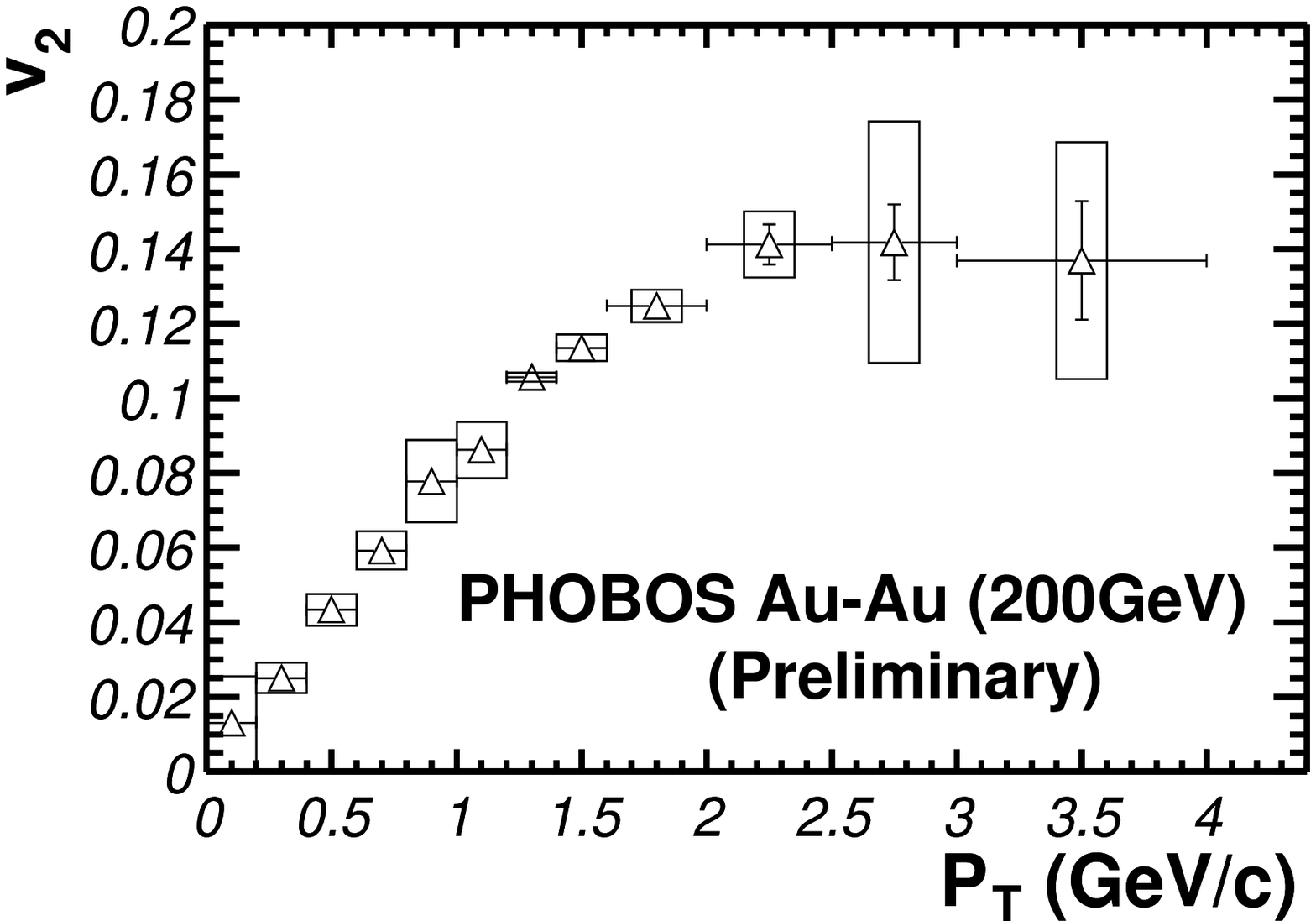,width=8cm} }
\caption{Elliptic flow as a function of transverse momentum for
Au-Au collisions at $\sqrt{s_{_{NN}}}$ = 200 GeV.} \label{v2pt}
\end{minipage}
\end{figure}

Results from this procedure are shown in Figures~\ref{v2CenHitTrk}
and~\ref{v2pt}.  The graphical representation of the errors in
these figures is similar to that for the first two figures.
Figure~\ref{v2CenHitTrk} shows the elliptic flow signal as a
function of the number of participants for the track-based
analysis overlayed with that from the hit-based analysis for 200
GeV data. The two techniques agree very well. This is significant
because of the differing sensitivity to background and non-flow
correlations. Figure~\ref{v2pt} gives the transverse momentum
dependence of the elliptic flow of charged hadrons in the 200 GeV
data.  The saturation observed at a transverse momentum greater
than 2 GeV/c is similar to what was observed at 130 GeV, and has
been interpreted as evidence for partonic energy loss through
gluon radiation in a dense system \cite{star_jq}.

\section{Bose-Einstein correlations}

Pairs of identified same-sign pions were used to calculate the
two-particle correlation functions.
The results were corrected for the effects of the tracking
algorithm and the Coulomb repulsion of the pions.  The
3-dimensional analysis of the correlation function using the
Bertsch-Pratt parameterization was performed in the LCMS frame in
the region of acceptance $0.2<y<1.5$ and $150<{\rm k_{T}}<350$
MeV/c. For the 15\% most central events, the preliminary fitted
source parameters for $\pi^{-}\pi^{-}$ ($\pi^{+}\pi^{+}$) pairs
are as follows: $\lambda = 0.54 \pm 0.02 (0.57 \pm 0.03)$,
R$_{out} = 5.8 \pm 0.2 (5.8 \pm 0.2)$ fm, R$_{side} = 5.1 \pm 0.4
(4.9 \pm 0.4)$ fm, R$_{long} = 6.8 \pm 0.3 (7.3 \pm 0.3)$ fm, and
R$_{out-long} = 4.9 \pm 1.7 (4.5 \pm 1.9)$ fm. The errors listed
are statistical only. In addition, there are systematic errors of
$\pm 0.06$ on the values of $\lambda$ and $\pm 1$ fm on the radii.
The results reported here are similar to those observed in Au-Au
collisions at $\sqrt{s_{_{NN}}} =$ 130 GeV
\cite{STARHBT,phenixHBT}.




\section{Summary}

Recent PHOBOS measurements of elliptic flow and two-particle
correlations in Au-Au collisions at $\sqrt{s_{_{NN}}} =$ 200 GeV
are very similar to values observed at $\sqrt{s_{_{NN}}} =$ 130
GeV. Notably, the new results at 200 GeV clearly show a saturation
of v$_{2}$ for p$_{T}>2$ GeV/c and a dramatic drop of v$_{2}$ as a
function of $|\eta|$.


\mbox{  }

{\footnotesize {\bf Acknowledgments:}  This work was partially
supported by US DoE grants DE-AC02-98CH10886, DE-FG02-93ER40802,
DE-FC02-94ER40818, DE-FG02-94ER40865, DE-FG02-99ER41099,
W-31-109-ENG-38, and NSF grants 9603486, 9722606 and 0072204. The
Polish group was partially supported by KBN grant 2-P03B-10323.
The NCU group was partially supported by NSC of Taiwan under
contract NSC 89-2112-M-008-024.}

\end{document}